\documentclass{ecai} 

\usepackage{latexsym}
\usepackage{amssymb}
\usepackage{amsmath}
\usepackage{amsthm}
\usepackage{booktabs}
\usepackage{enumitem}
\usepackage{graphicx}
\usepackage{color}
\usepackage{tikz}
\usepackage{subcaption}
\usetikzlibrary{positioning, arrows.meta, shapes.geometric}

\begin{document}

\begin{frontmatter}

\title{TABL-ABM: A Hybrid Framework for Synthetic LOB Generation}
\paperid{}
\author[1]{\fnms{Ollie}~\snm{Olby}}
\author[1]{\fnms{Rory}~\snm{Baggott}} 
\author[1]{\fnms{Namid}~\snm{Stillman}\thanks{Email: namid@simudyne.com}}

\address[1]{Simudyne\\London, UK}

\begin{abstract}
The recent application of deep learning models to financial trading has heightened the need for high fidelity financial time series data. This synthetic data can be used to supplement historical data to train large trading models. The state-of-the-art models for the generative application often rely on huge amounts of historical data and large, complicated models. These models range from autoregressive and diffusion-based models through to architecturally simpler models such as the temporal-attention bilinear layer. Agent-based approaches to modelling limit order book dynamics can also recreate trading activity through mechanistic models of trader behaviours. In this work, we demonstrate how a popular agent-based framework for simulating intraday trading activity, the Chiarella model, can be combined with one of the most performant deep learning models for forecasting multi-variate time series, the TABL model. This forecasting model is coupled to a simulation of a matching engine with a novel method for simulating deleted order flow. Our simulator gives us the ability to test the generative abilities of the forecasting model using stylised facts. Our results show that this methodology generates realistic price dynamics however, when analysing deeper, parts of the markets microstructure are not accurately recreated, highlighting the necessity for including more sophisticated agent behaviors into the modeling framework to help account for tail events.

\end{abstract}

\end{frontmatter}
\vspace{10pt}

\textbf{Keywords:} limit order book model, hybrid model, agent-based model, synthetic order flow

\section{Introduction}
Financial time series forecasting has had recent boosts in accuracy driven by improvements in deep learning, availability of larger datasets and more powerful compute resources \cite{ZhangAdvancements2024, kurisinkel2024text2timeseriesenhancingfinancialforecasting}. Financial time series forecasting is a valuable tool, where accurate predictions of price movements can have significant financial implications \cite{HAO2021106806}.

Limit Order Books (LOBs) are the primary way in which traders interact with the market \cite{LOB}. LOBs are queues of limit orders to be executed. The prioritization of the queue is determined by the exchange's protocols. For example, a common protocol is the price-time priority system \cite{angel1998priority}, where limit orders with the most favorable prices are prioritized, and orders at the same price level are ranked by their time of placement, with earlier orders receiving higher priority. Limit orders are important to traders as it ensures that the trader has control over the price they pay for an asset, as opposed to market orders where the price is determined by the current market condition.

Most forecasting models focus on predicting subsequent mid-prices, but to fully capture LOB dynamics, the next order or event must be predicted. This is because limit orders often arrive away from the best bid or ask, contributing to deeper liquidity. Reproducing the order flow is therefore more essential than simply forecasting the next mid-price for generating LOB dynamics. A key aspect of order flow forecasting, involves predicting various features beyond price, such as the size, and rate of orders. To do this, we built a multi-variate forecasting model which predicts the state of the next order in the sequence.

One application for this technology would be to allow funds to rehearse their execution strategies on simulated data. Data generated from rare events such as the 2008 financial crash or COVID-19 pandemic could be used to train deep learning trading models to perform under these extreme conditions. These generative models will also be reactive to trades made by a trader and so can be used to predict the market impact of an execution strategy.

Alongside deep learning methods of predicting LOB dynamics, agent-based modeling frameworks have also been employed for generating synthetic high frequency LOB data. One such framework is the extended Chiarella model \cite{Chiarella_2009}. It models traders as one of three types: fundamentalists, chartists, and noise traders. These traders can be calibrated to replicate empirical distributions, enabling realistic market behaviours \cite{Gao_2023}.

\begin{figure}[t]
    \centering
    \includegraphics[width=\linewidth]{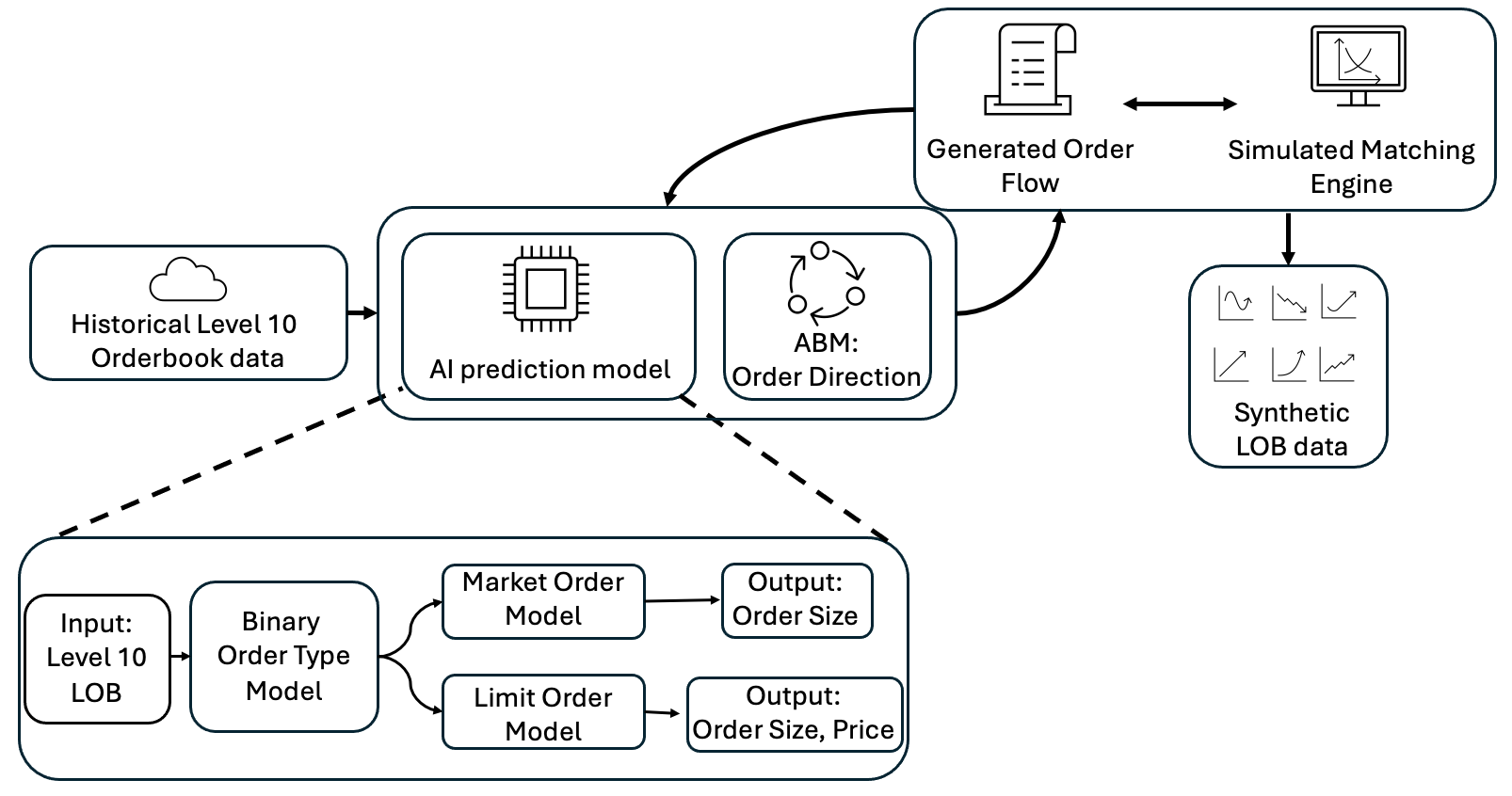}
    \vspace{0.3cm}
    \caption{High level diagram showing how deep learning and agent-based modeling work together to produce synthetic LOB data.}
    \vspace{0.75cm}
    \label{fig:hl_model}
\end{figure}

This paper proposes a novel strategy that combines state-of-the-art deep learning models and agent-based models to create synthetic LOB data. The model design is outlined in Figure~\ref{fig:hl_model}.

\section{Related Work}
Generating synthetic financial data is a large active area of research where the central challenge is to create synthetic data with sufficient accuracy for practical applications.
Many of the most performant deep learning models at this task are auto-regressive models, such as recurrent neural networks (RNNs), long short-term memory networks (LSTM) \cite{fjellström2022longshorttermmemoryneural}, and gated recurrent unit networks (GRUs), among others  \cite{NN4timeseries}. For a comprehensive review of deep learning models applied to forecasting time series, see \cite{prata2023lobbaseddeeplearningmodels}. While these models have shown strong empirical performance, they are often criticized for their lack of interpretability and are frequently treated as "black boxes", limiting their practical use in high-stakes financial decision-making.

One of the most impressive architectures of recent years is the transformer model and recent research has considered how these will perform at financial time series forecasting \cite{nagy2023generativeaiendtoendlimit}. Zeng et al. proposed a hybrid CNN-transformer model to help capture both the short and the long-term dependencies of the limit order book (LOB) \cite{zeng2023financialtimeseriesforecasting}. Alternatively, Emami et al. proposed a modality-aware transformer for financial time series prediction. This allows the model to incorporate both feature-level attention and temporal attention \cite{emami2024modalityawaretransformerfinancialtime}. This modality-aware model is especially promising in multi-modal data settings. Costa et al. demonstrated that transformer models can outperform LSTMs for financial time series forecasting \cite{Costa2023PredictionOS}, due to transformers’ superior memory capacity.

Forecasting models focus on trend prediction and lack the ability to generate the full multivariate structure of the LOB. They typically predict price movements, rather than the underlying order flow or market state transitions. One model suggested by Dong et al. aims to encode the current state of a time series into tokens \cite{dong2024metadatamatterstimeseries}. This leads to a more accurate state generation. Nagy et al. developed an autoregressive model using microstructure tokenisation for LOB prediction \cite{nagy2023generativeaiendtoendlimit}, showing promising scalability with intraday data. Huang et al. proposed a diffusion model for capturing the dynamics of the market state \cite{huang2024controllablefinancialmarketgeneration}, incorporating a meta-agent to generate realistic order flows. Similarly, Li et al. introduced the Large Market Model, which supports controllable order generation and is designed as a comprehensive market simulation and agent-training environment \cite{li2024marsfinancialmarketsimulation}.

A key family of models that bridge forecasting and efficiency are the Temporal Attention Bilinear Layer (TABL) models introduced by Tran et al. \cite{Tran_2019}, and later extended with bilinear normalisation \cite{tran2020datanormalizationbilinearstructures}. TABL models offer state-of-the-art results with simple architectures that remain computationally efficient.

In this work, we take the TABL approach to forecasting prices and integrate it with an agent-based model to generate synthetic LOB data. In the process, we demonstrate that the combination of relatively simple deep learning models and traditional simulation techniques, can achieve high fidelity data generation.

\section{Background}
\subsection{Limit Order Books}
Limit order books (LOBs) are central to the trading world as they allow for the prioritisation and matching of orders to buy or sell a specific volume of an asset. LOBs are a queue of all the limit orders that have been placed on a particular asset, the orders which are most favourable are called the best bid and best ask price and are placed at the front of the queue, while the difference between these best orders is known as the bid/ask spread. The bid/ask spread is a commonly-used measure of liquidity in financial markets, where it reflects the ease of trading and can serve as a proxy for market stability \cite{nelson2007selected}. A higher spread implies a large disparity between the buy and sell prices, which indicates an unstable market, whereas a smaller spread implies relative agreement about price, and a stable market. Limit orders are orders that are placed at a specific price and volume, such that the trader can be in control of the price point they want to trade at. Limit orders are characterised by price, volume and direction (bid/ask). Market orders are orders which effectively have a duration of zero, and don't ever populate the LOB. These are categorised by volume and direction only, as they are always placed at best ask and best bid price. Limit orders that cross the spread are treated as market orders because they match existing orders and execute immediately, without entering the LOB.


Orders in the LOB are normally matched via a price-time priority (although this is dependent on which exchange is being used). This means that price is the main priority, i.e. the better priced orders will be matched first. If two orders are placed at the same price, then the priority of matching comes down to who placed the order first.


\subsection{Stylised Facts}

Stylised Facts, are a collection of features which have been shown to exist in most financial time series \cite{vyetrenko2019realrealismmetricsrobust}. They are used as metrics to assess the fidelity of the simulation. The main facts analysed in this paper are:

\begin{itemize}
    \item \textbf{Autocorrelation of returns} Autocorrelation is a measure of how similar a series is with its lagged version. It is observed that the autocorrelation of returns is weak. This is because returns do not exhibit strong memory (past price movements do not dictate future ones).
    \item \textbf{Buy/sell autocorrelation} Buy/sell autocorrelation is observed to decay as a power law, due to the long-term persistence in the direction of orders.
    \item \textbf{Volatility clustering} Volatility clustering is measured by examining the autocorrelation of absolute returns. Financial returns exhibit volatility clustering, where periods of high volatility will be followed by high volatility and vice versa. This means we should observe a slow decay in the autocorrelation of the absolute returns.
    \item \textbf{Fat tailed distribution} The return distributions in financial time series are characterized by fat tails. Therefore, the series should exhibit a high level of kurtosis. The extent of accuracy for this stylised fact can be determined by comparing the Hill index for the distribution of returns.
\end{itemize}

\subsection{Agent-Based Model}

The extended Chiarella model is an agent-based modelling framework for modeling market dynamics \cite{Chiarella_2009}. There are three types of agents considered: fundamental, momentum, and noise traders. 

\begin{itemize}
    \item The \textbf{fundamental trader} makes trading decisions based on the price differential between the mid-price (\(p_t\)) and the fundamental price (\(v_t\)).
    \item The \textbf{momentum trader} bases their decisions on current market trends, such as the direction and timing of price movements, aiming to take advantage of these trends.
    \item The \textbf{noise trader} generates uncorrelated trading decisions and reflects both market features that aren't captured by this simplified model as well as uncertainty within the market.
\end{itemize}

Each trader's behaviour is governed by a simple equation. The fundamentalist is characterised as
\begin{equation}
    D_{\text{fundamental}} = \kappa \left( v_t - p_t \right)
\end{equation}
where $\kappa$ is a calibrated parameter. The fundamental price of the asset ($v_t$) is modelled as a Geometric Brownian Motion where the mean and standard deviation are fitted to the empirical price path.

The momentum trader places orders according to the strength of the momentum signal
\begin{equation}
    M_t = (1 - \alpha) M_{t-1} + \alpha (p_t - p_{t-1})
\end{equation}
where $\alpha$ is the decay rate and $M_t$ is the momentum signal. From this the demand from the momentum trader can be found
\begin{equation}
    D_{\text{momentum}} = \beta \tanh \left( \gamma M_t \right)
\end{equation}
where $D_{\text{momentum}}$ is the demand from the momentum trader, $\beta$ is a parameter that determines the strength of the momentum trader and $\gamma$ is a parameter which describes the saturation of the momentum signal and is calibrated.

Finally, the noise trader is modelled by a simple Gaussian distribution. The noise trader is shown by
\begin{equation}
    D_{\text{noise}} \sim \mathcal{N}(0, \sigma_N)
\end{equation}
where $D_{\text{noise}}$ is the demand from the noise traders, $\sigma_N$ is the standard deviation and $\mathcal{N}$ denotes a normal distribution.

The overall demand can be given as:
\begin{equation} \label{eq:chiarella}
    D_{\text{overall}} = \left( D_f \Delta T + D_m + D_{\text{noise}} \sqrt{\Delta T} \right)
\end{equation}
where $\Delta T$ is the change in time between orders.

\section{TABL-ABM}
To accurately recreate the order flow, three architecturally similar models were developed. The first model was designed as a binary classifier to predict the type of order, either a limit order or a market order. Based on this prediction, the logic then branches to either the limit order model or the market order model. The limit order model predicts both the size and price of the order, while the market order model predicts only the size, this is shown in Figure~\ref{fig:hl_model}.

These three models are designed as an extension of the work done by Tran et al. on the Temporal Attention Augmented Bilinear Layer (TABL) framework \cite{Tran_2019} which has been shown by a recent benchmarking study to have state-of-the-art accuracy and computational efficiency when compared with other models \cite{prata2023lobbaseddeeplearningmodels}. The TABL is extended to perform a multivariate prediction task alongside an ABM.

The TABL model combines bilinear layers and an attention mask to help capture relevant dynamics for LOB forecasting. In the next section we outline how this forecasting model can be used for synthetic LOB generation.

\subsection{Model Layers}
Bilinear layers are a special type of neural network (NN), that takes two different input vectors and learns the interactions between the two to generate a suitable output. This lets the model capture interactions between two dimensions at the same time. In our example, this facilitates the model to learn the interactions between both the temporal and feature dimensions simultaneously. This is particularly beneficial for financial data, where modeling the way the features interact with the temporal dimension is useful for generating accurate predictions.
Bilinear layers are built using linear transformations and are easier to analyse than standard multi-layer perceptron (MLPs). This drop in relative complexity does not equate to a drop in performance, and bilinear layers will often outperform an MLP of the same size \cite{sharkey2023technicalnotebilinearlayers}.
 
A bilinear layer learns interactions between two input dimensions. They multiply two input vectors and learn a weight for each pairwise interaction. The mapping for which is expressed as
\begin{equation}\label{eq:Bilinear_Layer}
    Y = \phi (W_1 X W_2 + B)
\end{equation}
where \( Y \) is the output of the bilinear layer, \( \phi \) is a non-linear activation function such as ReLU, sigmoid, or tanh, \( X \) is the input, $W_1$ and $W_2$ are learnable weight matrices and \( B \) is a learnable bias matrix.

To aid interpretability and performance the TABL model uses an attention layer. The attention layer assigns varying importance to different time instances to determine which points in time are most relevant for the prediction task. It uses a learned matrix to weight the temporal dependencies which is applied to an attention mask that emphasises the important time steps.

\subsection{TABL Model}
\begin{figure}[t]
    \centering
    \includegraphics[width=0.8\linewidth]{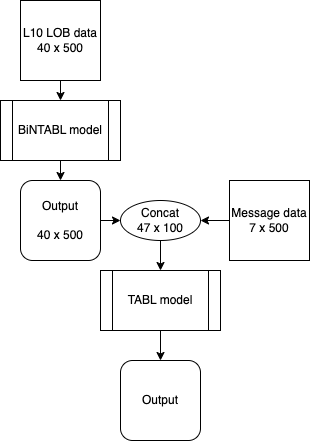}
    \vspace{0.3cm}
    \caption{Model architecture schematic. LOB data is processed by a BiNTABL model, and its output is concatenated with message data before being passed to a TABL model for final prediction.}
    \label{fig:whole_model}
    \vspace{0.75cm}
\end{figure}

In order to build a realistic model of order flow, we combine a bilinear normalisation TABL (BiNTABL) module and a TABL module. The BiNTABL module processes the book data to extract insights into the microstructure of the LOB. These insights are then concatenated with the order flow from the message data and passed through the TABL module to produce the next order state. The BiN layer performs normalisation along both the temporal and feature dimensions, allowing the model to generalise better and also standardise the input data. The whole model layout is shown in Figure~\ref{fig:whole_model} and the architecture of the model can be seen in the paper by Tran et al. \cite{tran2020datanormalizationbilinearstructures}.

\subsection{Agent-Based Model}
The Chiarella model is designed to capture the overall demand of traders by incorporating momentum, fundamental value, and noise-driven behaviours, which are common components of real-world trading. In this framework, the sign of the demand reflects the likely direction of the next order: a positive demand suggests a buy order, while a negative demand implies a sell. By leveraging this structure, we use the Chiarella model to statistically infer the direction of the next order under the assumption that traders act according to these dynamics. This framework also anchors the model to a fundamental price, thereby preventing it from diverging into out-of-distribution values that the TABL architecture may not be well-equipped to handle.

This separation of roles within the model allows us to combine the strengths of both behavioural (agent-based) finance and machine learning. The Chiarella model introduces a framework that is more grounded in financial behaviours than purely data-driven AI models. In contrast, the machine learning model is tasked with generating the price, size, and type (market vs. limit). These characteristics are often shaped by short-term market microstructure, for which data-driven approaches are highly effective.

By combining deep learning and ABM, we enable better predictive abilities when the generated paths deviate from the historical data. This is especially key when building a model using a small amount of data.

\section{Simulated Matching Engine} \label{sec:sme}
We propose a simulated matching engine designed to perform two tasks. The first task is to correctly queue limit orders and match any market orders. The second task is to model the deletion of orders. This is used to assess the fidelity of the deep learning model in its ability to create realistic order flow. There are several limitations to the matching engine. First, it operates in event time rather than clock time, meaning it does not model real-world temporal intervals between events. Second, in rare instances where multiple orders are deleted simultaneously or a large market order results in the execution of many resting orders, the matching engine imposes constraints on the order book depth. It maintains a maximum depth of 25 levels and a minimum depth of 10 levels. When the depth falls below this threshold, noise orders are introduced to restore the depth, positioned between 5 and 10 ticks away from the furthest existing price level.

\subsection{Deleted Orders} \label{sec:deleted_orders}
To reproduce deleted orders, a per time step probability of deletion was determined per order. This probability is calculated based on the orders' depth of insertion, its current depth in the limit order book and its current duration. We determine the probability that an order is at a certain depth, given that it is deleted, by

\begin{equation}
P(\text{Depth} = x \mid \text{Deleted}=True) = \frac{P(A \cap B)}{P(B)} = \frac{N^x_d}{N_d}
\end{equation}
where, event $N^d_x$ is the number of orders deleted from a certain depth and $N_d$ is the total number of limit orders deleted. For ease of notation, $A$ will represent the event: $\text{Deleted} = \text{True}$ and $B$ will represent $\text{Depth} = x$. Using Bayes theorem, it is possible to find the probability that an order is deleted given its current depth.

\begin{equation}
    P(A \mid B) = \frac{P(B \mid A) P(A)}{P(B)}
\end{equation}
\begin{equation}
    P(A \mid B) = \frac{P(B \mid A) \times \frac{N_d}{N_t}}{\frac{N_p^x}{N_t}}
\end{equation}
where $N_p^x$ is the number of limit orders placed at a depth of $x$ and $N_t$ is the total number of limit orders placed. The assumption $P(\text{Depth} = x) = \frac{N_p^x}{N_t}$ is made, as this serves as a measure of relative activity at each depth. A more precise metric would involve tracking individual orders and determining the probability of their progression to a specific depth. This is very computationally heavy and involves in depth LOB data, hence we use this simplifying assumption.

Given the probability that an order will be deleted for its current depth, the per time step probability needed to be found. The probability of an order being deleted can be represented as a conditional probability tree with $T$ time steps, where the probability of deletion at each event step is $P_e$. The probability can be calculated as the complement of the probability of the order not being deleted throughout all $T$ steps.

\begin{equation}
    P(A\mid B) = 1-(1-P_e)^T
\end{equation}
The per time step probability $P_e$ is then
\begin{equation} \label{eq:DelProb}
    P_e = 1-(1-P(A \mid B))^{1/(T_0-T_t)}
\end{equation}

The total duration, $T_0$, of the order is estimated empirically from its insertion depth. Then, as the order moves through the LOB the probability changes according to equation~\ref{eq:DelProb}. After each event all probabilities are updated to reflect the current duration of each order in events, $T_t$. The probabilities are then given a small scaling factor to help match the empirical results directly.

Hence, the duration of an order is a function of the depth at which it is placed, its current depth in the LOB and how long it has been in the LOB (current duration), given by
\begin{equation}
    P(\text{Deletion}) = f(\text{Depth}_i, \text{Depth}_t, T_t)
\end{equation}
where $\text{Depth}_i$ is the insertion depth, $\text{Depth}_t$ is the current depth and $T_t$ is how long the order has been in the LOB in time steps.
This value is then used to calculate the per event probability ($P_e$) for each event step. This calculation is repeated at each event, as $P(\text{Deleted} = \text{True} \mid \text{Depth} = x)$ changes dynamically with the current position of the order ($x$) in the LOB and the current duration for each order ($T_t$). The per event probability, $P_t$, is evaluated at each event. A random number is drawn from a uniform distribution over [0,1], and the order is deleted if the random number is less than $P_t$.

\section{Methodology}
In this section we explain how the data and TABL-ABM model is set up and trained before being tested in the simulator to assess its performance in replicating stylised facts.

\subsection{Data and Preprocessing}
In this work, we use the LOBSTER sample limit order book data \cite{huang2011lobster}. The data consists of both order book files and message files for 5 different symbols (MSFT, AAPL, GOOG, INTC, AMZN). The message data is in the form: time stamp, event type, order ID, size, price, direction. For this project we used the level 10 orderbook data from Apple (AAPL).

Market orders are not shown in the messages of the LOB as the orders never enter the LOB. Therefore, they have to be inferred from the execution of limit orders on the opposite side of the book. We can be reasonably confident about aggregating consecutive executions of limit orders into larger market orders as the executions occur at the same time stamp and have the same direction.

An additional feature added to the message data is the signed contributions to the limit order book. It is defined as \cite{Cont_2013}
\begin{align}\label{eq:OFI}
e_n = & \ q_b^n \mathbb{I}\{P_b^n \geq P_b^{n-1}\} 
      - q_b^{n-1} \mathbb{I}\{P_b^n \leq P_b^{n-1}\} \nonumber \\
      & - q_s^n \mathbb{I}\{P_s^n \leq P_s^{n-1}\} 
      + q_s^{n-1} \mathbb{I}\{P_s^n \geq P_s^{n-1}\}
\end{align}
where $q$ is the volume and $P$ is the price. This feature, denoted as $e_n$, is a measure of order flow imbalance (OFI) at a per order resolution, by accounting for both direction and magnitude of changes to bid/ask prices and volumes. It provides the model with insight into the current microstructure of the LOB, this insight helps the model to predict the state of the next order \cite{anantha2024forecastinghighfrequencyorder}.

The data is divided into three distinct subsets. The first includes all orders, combining both limit and market orders. The second contains only market orders, and the third consists solely of limit orders. Three separate models are then trained using the same architecture shown in Figure~\ref{fig:whole_model}. The first model, trained on the full dataset, is a binary classifier that predicts the order type (market or limit). The second model, trained on the market order subset, focuses on market orders and predicts the order size. The third model, trained on the limit order subset, handles limit orders and is a multi-class classifier that predicts both the size and price level of the order.

The labels used for each subset are defined as follows. For the full dataset, the label is binary, 1 for market orders and 0 for limit orders. In the limit order subset, the model predicts two targets: order size and price distance, where price distance refers to the difference between the order’s price and the current best bid or ask (i.e., the touch line). Both features are discretised into classes: size is divided into 20 bins, and price distance into 40 bins. These bins are constructed to cover 80\% of the data for size and 90\% for price distance, ensuring the model focuses on the most frequently occurring values. The bins are not evenly spaced, they are designed to reflect the non-uniformity of financial markets. For the market order subset, the label is limited to order size, and the same 20-class binning strategy used for limit orders is applied.

Finally, each subset of data was reshaped so that each instance contained the last 500 events. This was then resampled into training, validation, and testing datasets in the following percentages: 64\%, 16\% and 20\% respectively. A min/max scaler was applied to the data to help regularise the weights during training.

\begin{figure}[t]
    \centering
    \includegraphics[width=\linewidth]{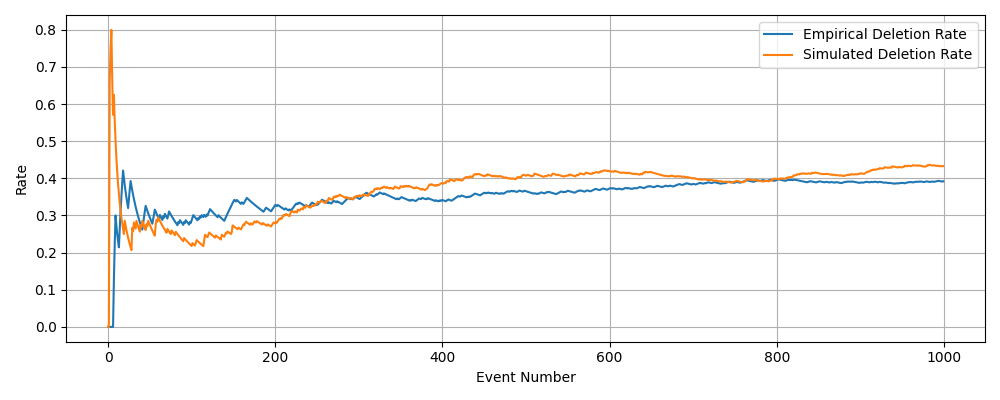}
    \caption{Cumulative order deletion rate over simulated and historical events.}
    \label{fig:deletions}
    \vspace{0.5cm}
\end{figure}

\subsection{Training and Model Parameter Set Up}

The model uses a learning rate scheduler which reduces the learning rate by a factor of 0.1 when the model has plateaued for 5 epochs. Checkpointing and early stopping are used, such that if the validation loss does not improve for 10 epochs, training is halted and the model reverts to the version with the lowest recorded validation loss. A dropout strength of 0.3 was included to help reduce the effects of overfitting. For the limit and market order models, the loss function used was a focal loss with a gamma of 2, this was chosen to help limit the effects of class imbalance. The order type model was formulated as a binary classification task, with a positive class weight of 2 to address class imbalance.

\subsection{ABM Calibration}
The extended Chiarella model required calibrating to the asset and day in question. The fundamental value ($v_t$) is modeled as a Geometric Brownian Motion with $\mu$ equal to the average historical price and $\sigma$ is equal to the average volatility of the historical data. The decay rate for the momentum traders $\alpha$ is fixed as $\alpha = 1/(1+\tau)$ where $\tau$ is equal to the look back period for the momentum signal \cite{majewski2018coexistencetrendvaluefinancial}. Here, $\alpha$ was fixed as $\alpha = 1/(1+h)$, where $h$ is the typical horizon of trend computation which is decided to be 1 day, which gives $\alpha = 0.5$ \cite{Gao_2023}. $\gamma$ is also fixed as $\gamma^{-1} = 2\sigma$ where $\sigma$ is the standard deviation of the momentum signal ($M_t$).

The following parameters were calibrated using a grid search to minimise a loss function based on the stylised facts ($\beta$, $\kappa$, $\sigma_N$, $\gamma$).

\begin{equation}
\begin{aligned}
\mathcal{L}(\theta) =\ & 
\left| H_{\text{sim}} - H_{\text{hist}} \right| 
+ \left| \sigma_{\text{sim}} - \sigma_{\text{hist}} \right| + \left| K_{\text{sim}} - K_{\text{hist}} \right|\\
&+ \sum_{i=1}^{9} \left| \rho^{(r)}_{\text{sim}}(i) - \rho^{(r)}_{\text{hist}}(i) \right| \\
&+ \sum_{i=1}^{9} \left| \rho^{(r^2)}_{\text{sim}}(i) - \rho^{(r^2)}_{\text{hist}}(i) \right|
\end{aligned}
\label{eq:chiarella_loss}
\end{equation}
where $\mathcal{L}$ is the loss function, $H$ is the Hill index, $\sigma$ is the standard deviation of returns, $\rho^{(r)}(i)$ is the autocorrelation of returns at lag $i$, $\rho^{(r^2)}(i)$ is the autocorrelation of squared returns at lag $i$ and $K$ is the kurtosis of the signal.

Equation \ref{eq:chiarella} shows how these values can be converted into a demand. If the demand is positive the order will be a buy and if the demand is negative then the order is a sell.

\subsection{Simulated Matching Engine}

A matching engine is used to test the efficacy of the TABL-ABM model by using stylised facts as the main performance metric. The engine works in tandem with the TABL-ABM whereby the TABL-ABM model produces the next order in the sequence based on the current and previous state of the LOB. This order is then executed using the simulated matching engine. If the order is a market order then limit orders on the opposite side of the book are liquidated to match the volume of the market order. If the order is a limit order then the order will join the queue as per the LOB queuing protocol used (in this case a price-time priority). Once the new orders have been added, the matching engine has the opportunity to delete any orders based on empirical statistics as shown in Section~\ref{sec:deleted_orders}. The LOB is then updated and then passed to the TABL-ABM to get the next order in the sequence.

\begin{figure*}[!t]
    \centering
    \begin{subfigure}[b]{0.47\textwidth}
        \centering
        \includegraphics[width=\linewidth]{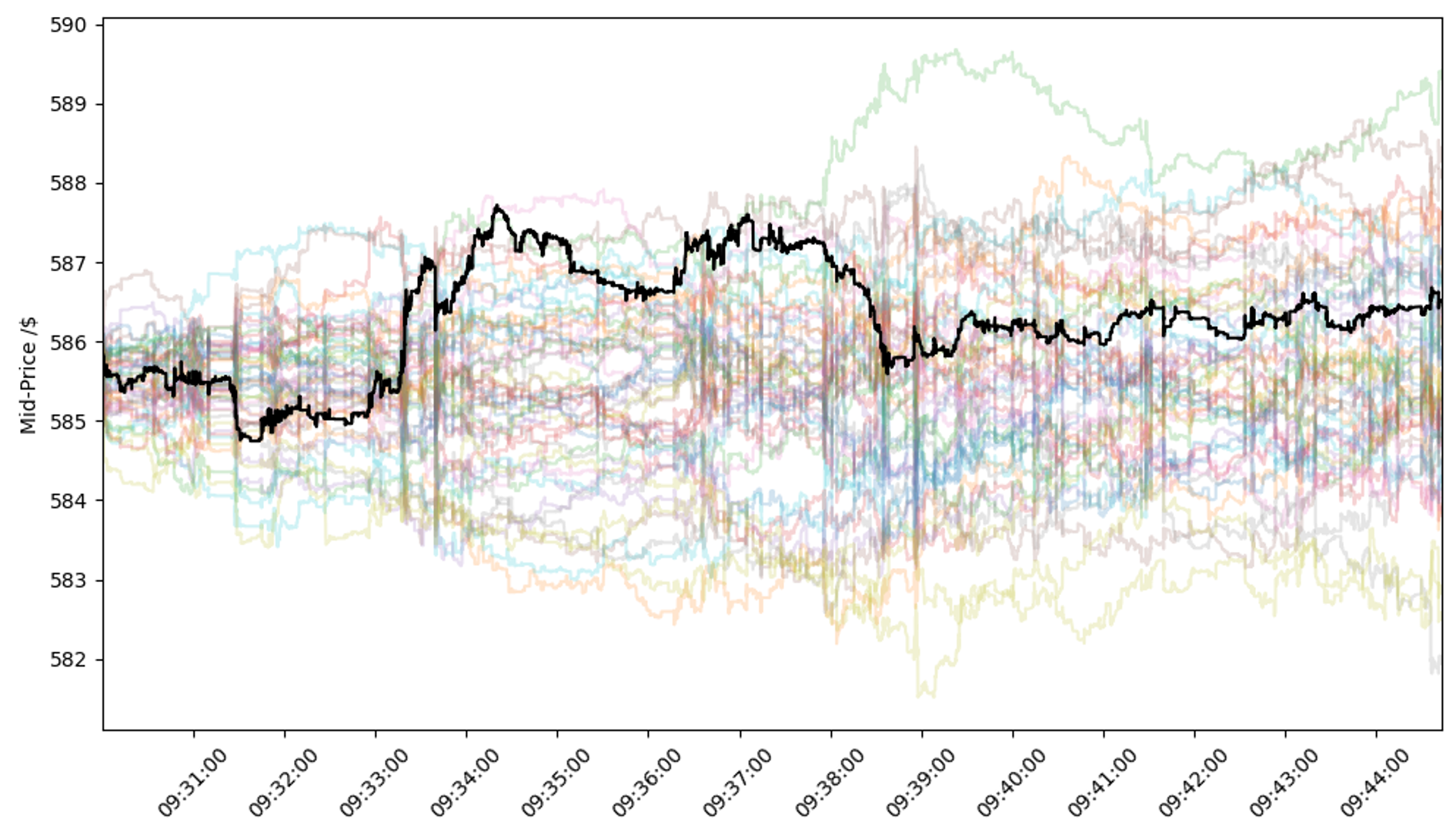}
        \caption{Historical and Monte Carlo simulation mid-price paths.}
        \label{fig:simulated_path}
        \vspace{0.5cm}
    \end{subfigure}
    \hfill
    \begin{subfigure}[b]{0.47\textwidth}
        \centering
        \includegraphics[width=\linewidth]{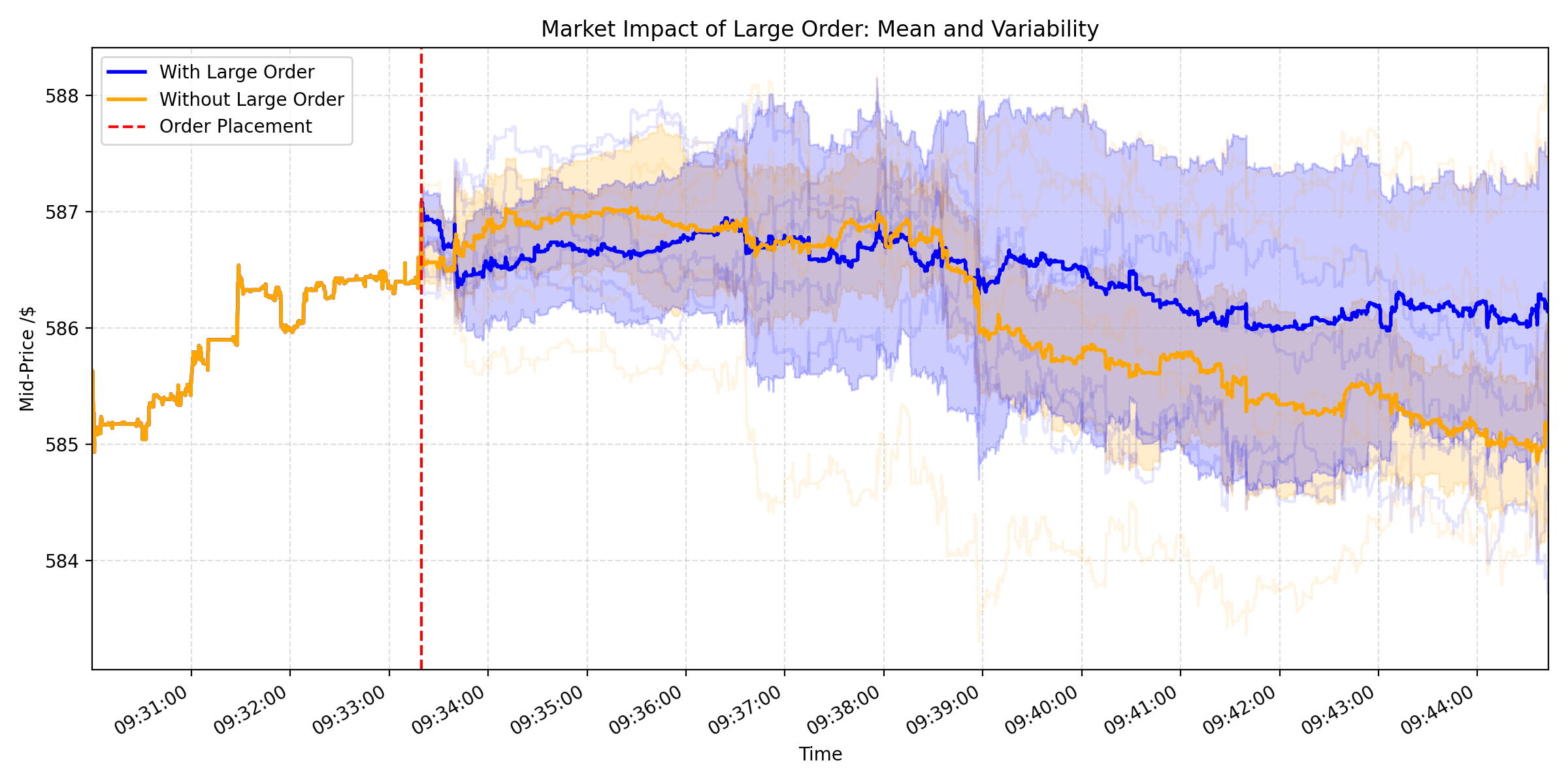}
        \caption{Demonstrating market impact}
        \label{fig:market-impact}
        \vspace{0.5cm} 
    \end{subfigure}
    \caption{Market simulation results showing (a) Monte Carlo simulations of mid-price paths compared to historical data,  where we display fifty simulated mid-price trajectories generated by the model, overlaid with the historical mid-price of Apple (21/6/2012). In (b), we show the market impact of a large order, shown by mid-price paths from fifty simulations, where twenty-five have a large market order and twenty-five stochastic realisations of the baseline scenario. A red dashed vertical line shows the moment at which a large order is introduced.}
    \vspace{0.5cm} 
    \label{fig:combined_market}
\end{figure*}
\section{Results and Discussion}





First, we assess the order deletion rate to check if our framework for modeling deleted orders, as proposed in this paper, is effective at reproducing empirical dynamics. Figure~\ref{fig:deletions} shows how the empirical and simulated order deletion rates change over time. It can be seen from both historical and simulated that the rate is more volatile at the start of the day, and that this volatility then subsides to where the deletion rate reaches a stable level of roughly 0.4 deletions per event. This plot highlights how the simulated order flow reproduces the historical deletion rates and supports our novel deletion methodology as a viable solution to introducing order deletions into simulated limit order book dynamics.

We next look into our models ability to reproduce realistic price paths. Figure~\ref{fig:simulated_path} shows the generated mid-price path for 50 Monte Carlo runs over 10,000 simulated events. These paths illustrate the model's ability to capture key features of the market, such as the volatility and mean-reverting behaviour. These realistic trajectories support the model's validity as a tool for simulating financial markets.

To further explore the usefulness of our model, we conduct a counterfactual market impact experiment, shown in Figure~\ref{fig:market-impact}. This figure highlights how the simulated market initially responds with a sharp price movement. This is followed by a gradual reversion toward the original price trajectory, indicating a degree of market resilience or recovery once the imbalance subsides. This behavior is consistent with empirical observations in financial markets. This supports the potential use of the model in stress testing and in the design of execution algorithms. 

\begin{figure*}[!t]
    \centering
    \begin{subfigure}[b]{0.47\textwidth}
        \centering
        \includegraphics[width=\linewidth]{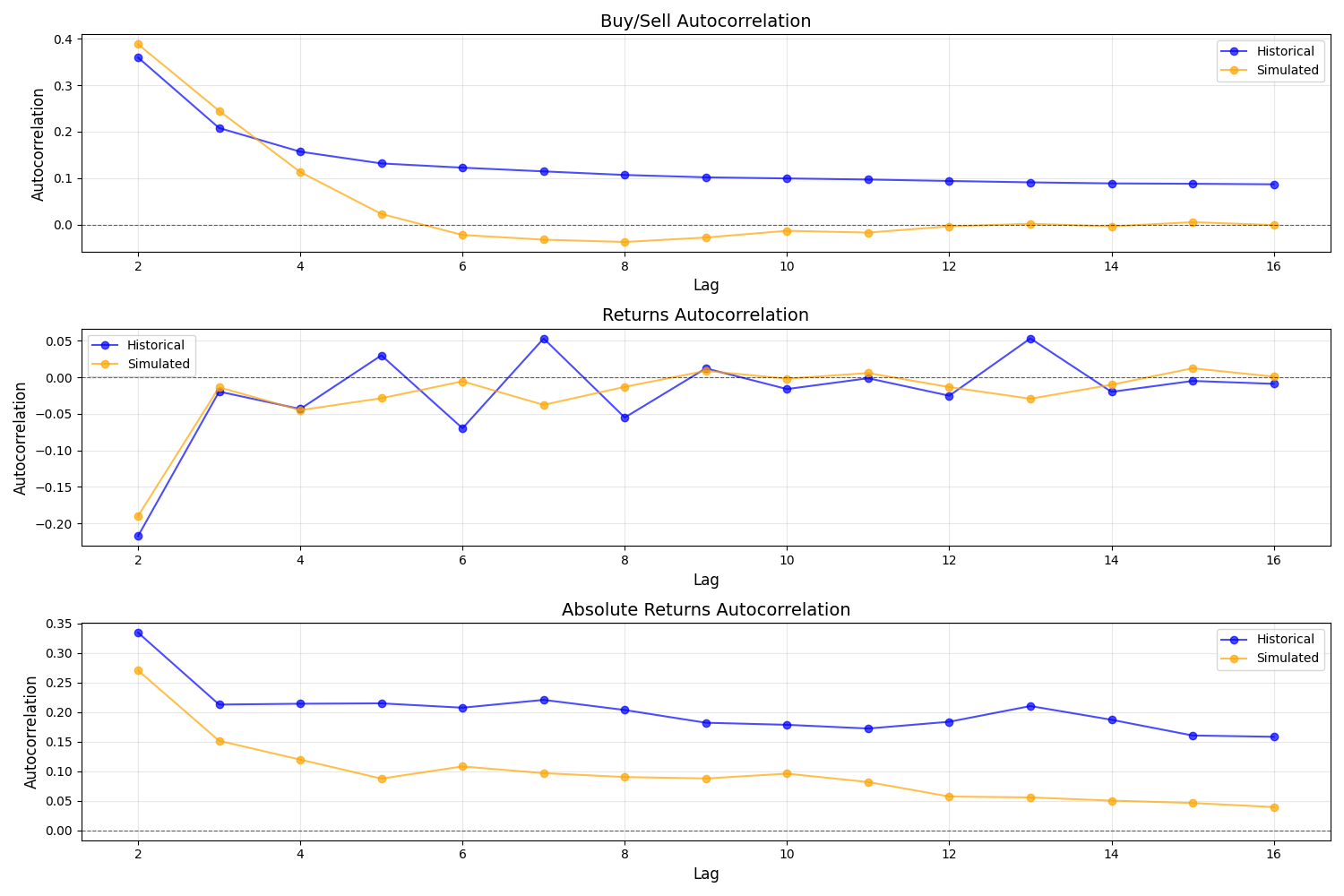}
        \caption{Comparison of autocorrelations.}
        \label{fig:autocorrelations}
        \vspace{0.5cm}
    \end{subfigure}
    \hfill
    \begin{subfigure}[b]{0.47\textwidth}
        \centering
        \includegraphics[width=\linewidth]{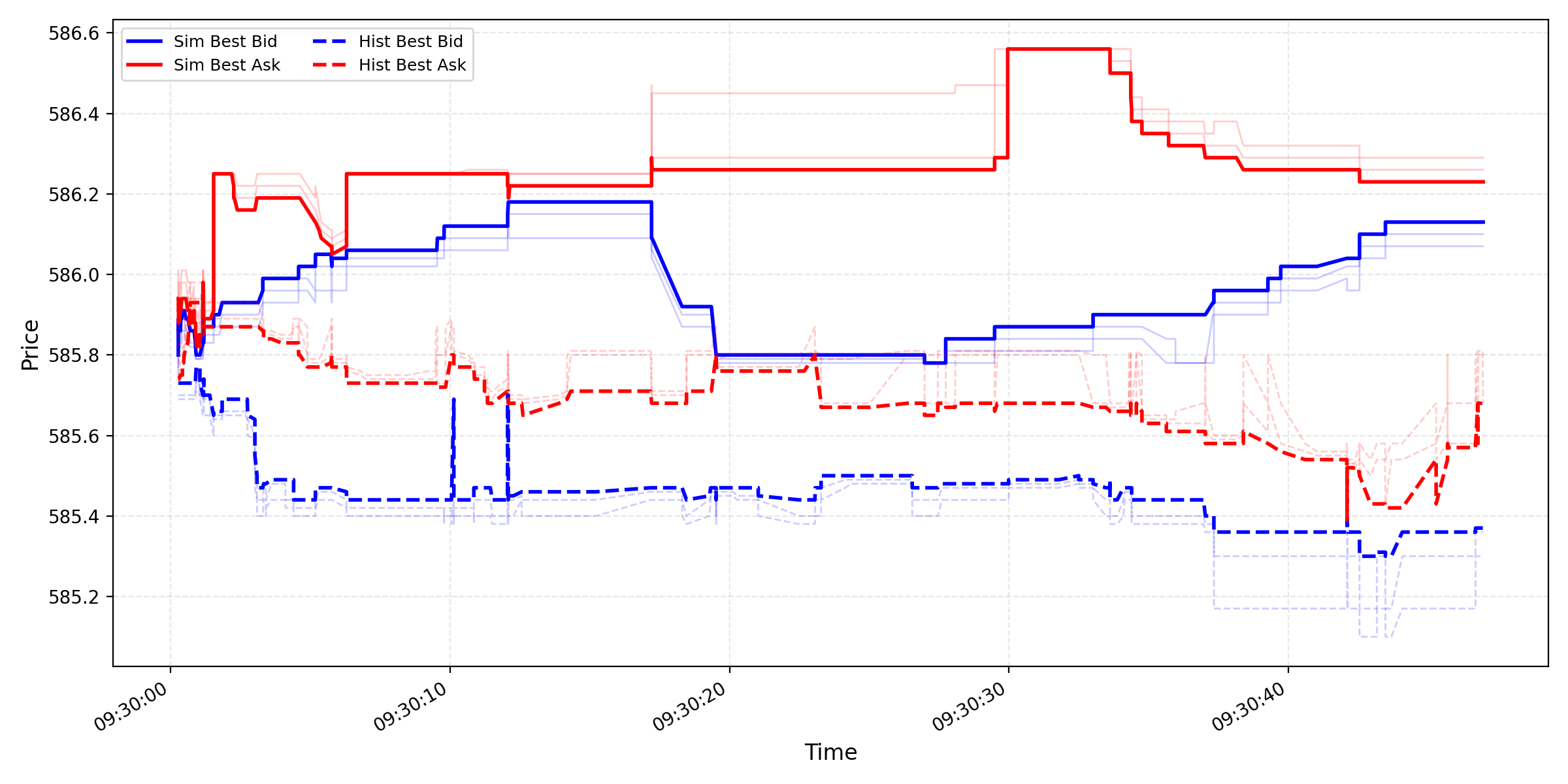}
        \caption{Comparison of order book levels.}
        \label{fig:levels}
        \vspace{0.5cm}
    \end{subfigure}
    \caption{Validation of simulation fidelity showing (a) autocorrelation between historical and simulated data for buy/sell indicators, returns, and absolute returns and (b) comparison of order book level dynamics between simulated and historical data, showing the top three bid and ask levels over time for both the historical and simulated environment. Best bid and ask prices are shown with stronger line weights, while deeper levels are faded for clarity. The plot begins shortly after the opening auction to exclude early-day volatility.}
    \label{fig:combined_validation}
    \vspace{0.5cm}
\end{figure*}

To assess the fidelity of the price paths, we compute several stylised facts, commonly observed in real markets. Figure~\ref{fig:autocorrelations} compares the autocorrelations of buy/sell indicators, returns, and absolute returns between the simulated and empirical datasets. 

The expected behaviour for the bid/ask indicators autocorrelation is that the correlation decreases as a power law due to the observed direction persistence in the market. The model uses the Chiarella model to determine the direction, due to this being influenced by noise traders, the model has less persistence of the direction of orders at higher lags. It is possible that this can be improved by calibrating the Chiarella model not for minimising Equation~\ref{eq:chiarella_loss} but for the order flow direction generation. 

It is expected that there is almost no autocorrelation with the returns. This is observed in the results from the model and the results are well matched to the historical levels. This shows that the model is aligning well with the efficient market hypothesis, meaning the model's prices are sufficiently unpredictable in the short term. The last autocorrelation analysed is the absolute returns autocorrelations. This is a measure of volatility clustering, we expect to see that there is a correlation in the absolute returns at longer lags. We observe a similar decay pattern to the empirical volatility clustering, however, we do not see similar levels to the historical at longer lags. This shows that our model does exhibit short term volatility memory but lacks persistence. Finally, the distributions of price fluctuations is examined using the Hill index, a statistical measure used to quantify the heaviness of the tails in a distribution. A smaller Hill index indicates fatter tails, implying a higher likelihood of extreme returns. The empirical Hill index of 0.27 points to a pronounced heavy-tailed behavior in the historical data. In contrast, the simulated mid‐price returns yield a Hill index of 0.49, indicating that the simulation produces lighter tails than those observed empirically.

These discrepancies prompted a closer investigation into the underlying microstructure of our model, which we examine in Figure~\ref{fig:levels}. This figure shows how the historic and simulated LOB dynamics differ and helps to see the limitations of the model. It is observed that the spread is wider and more diverse than the historic. However the key observation is that the spread and mid-price change more gradually in the simulator and the simulator is not prone to instantaneous jumps from either the best bid or best ask price. These jumps can be seen in Figure~\ref{fig:levels} where we see very aggressive, but likely small volume, limit orders being placed far across the spread, these are then quickly deleted or executed and the spread jumps back immediately. These events cause very fast and large changes to the mid-price. The events are not adequately recreated in the model and could explain why the simulation experiences a lighter tail than the historical data.

The lack of persistence in volatility clustering can also be explained by the microstructure deficiencies. Without the stabilising presence of liquidity providers, the spread is more volatile and less stable. This makes sustained volatility patterns observed in real markets difficult to reproduce. This unstable microstructure fails to generate the longer-term volatility memory that characterises empirical financial data.

Despite the promising alignment of some stylised facts, these metrics alone do not fully reflect the true quality or realism of the simulated market. While our model is successful in generating plausible price paths and shock reactions, closer inspection reveals structural differences in the price dynamics that a more comprehensive analysis of stylised facts would likely help to diagnose. This will be the topic of future work.

\section{Conclusion}
In this work, we have shown how the combination of the proposed TABL-ABM model and a simulated matching engine produces realistic, synthetic order flow. The realism of the data is validated by comparing the resulting mid-price path to established stylized facts. This is made possible through a novel agent-based modeling (ABM) approach for simulating order direction, along with the statistical modeling of order deletions within the simulator. Despite the qualitatively strong results and support for some stylised facts, some key parts of the market microstructure are not well replicated. This emphasises the need for more robust and comprehensive evaluation methods to assess the fidelity of synthetic LOB data, which will be the topic of future work.


\bibliography{bib}
\end{document}